\documentclass[twocolumn]{revtex4}
\usepackage{graphics,epsfig}

\def\lsim{\lower.5ex\hbox{$\; \buildrel < \over \sim \;$}}
\def\gsim{\lower.5ex\hbox{$\; \buildrel > \over \sim \;$}}
\def\g{\ifmmode \gamma \else $\gamma$\fi}
\def\gs{\ifmmode \gamma \else $\gamma~$\fi}

\begin{document}

\title{Reconstructing $\rho^0$ and $\omega$ mesons from non-leptonic decays in
C+C at 2AGeV
}

\author{Sascha Vogel and  Marcus Bleicher}

\affiliation{\vspace*{3mm}Institut f\"ur Theoretische Physik, J.W. Goethe Universit\"at, \\
Max von Laue Stra\ss{}e 1,\\
60438 Frankfurt am Main, Germany
}
\maketitle

\noindent We predict transverse and longitudinal momentum spectra
and yields of $\rho^0$ and $\omega$ mesons reconstructed from hadron
correlations in C+C reactions at 2~AGeV.  The rapidity and $p_T$
distributions for reconstructable $\rho^0$ mesons differs strongly
from the primary distribution, while the $\omega$'s distributions are
only weakly modified. We discuss the temporal and spatial
distributions of the particles emitted in the hadron channel.
Finally, we report on the mass shift of the $\rho^0$ due to its
coupling to the $N^*(1520)$, which is observable in both the
di-lepton and $\pi\pi$ channel. Our calculations can be tested with
the Hades experiment at GSI, Darmstadt.

\vspace{.6cm}


The Hades spectrometer at GSI opens the possibility to measure
di-lepton yields and spectra at moderate energies with unprecedented
accuracy. However, in addition it allows also to get complimentary information
on the most interesting di-lepton sources (the $\rho^0$ and $\omega$
mesons) directly from the reconstruction of their hadronic decay products. 
I.e. the direct reconstruction of the invariant mass spectra of the resonances 
from 2- and 3-particle pion correlations.
Over the last years the exploration of resonance yields and spectra from 
hadron correlation has attracted a great amount of experimental \cite{Markert:2004xx,Markert:2003rw,Markert:2002xi,Fachini:2003dx,Fachini:2003mc,Fachini:2004jx,Adler:2004zn, Adams:2004ep}
and  theoretical attention \cite{Johnson:1999fv,Soff:2000ae,Torrieri:2001ue,Torrieri:2001tg,Bleicher:2002dm,Torrieri:2002jp,Bleicher:2002rx,Bleicher:2003ij,Torrieri:2004zz,Vogel:2005qr}. 

Detailed experimental studies
of hadron resonances at the full SPS and highest RHIC energy
\cite{Markert:2004xx,Markert:2003rw,Markert:2002xi,Fachini:2003dx,Fachini:2003mc,Fachini:2004jx,Adler:2004zn, Adams:2004ep} have recently been performed. 
Theoretically there are still unsolved problems concerning the absorption and
regeneration of resonances after chemical decoupling of the system. It is
expected that the inclusion of pseudo-elastic interactions between chemical and 
kinetic freeze-out might solve some discrepancies of  statistical models 
for resonance yields at high energies \cite{Braun-Munzinger:1995bp, Braun-Munzinger:2001ip, kaischweda}.
At low energies, the first experimental investigations of resonances are
eagerly awaited to explore in-medium effects on vector mesons near 
nuclear groundstate densities in the 1-2 AGeV beam energy range.

Especially resonances that have electromagnetic and hadronic decay channels
allow deeper insights into the dynamics and lifetimes of the hot and dense hadronic
matter stage created in light and heavy ion reactions.
That is because the leptonic decay channel carries  
information from the early stages of the reaction, since the
leptons can leave the dense regions undisturbed,
while the hadronic decay channel triggers on the late stages (near kinetic decoupling) 
of the evolution.
\begin{figure}[hbt]
\vspace*{-1cm}\centerline{\epsfig {file=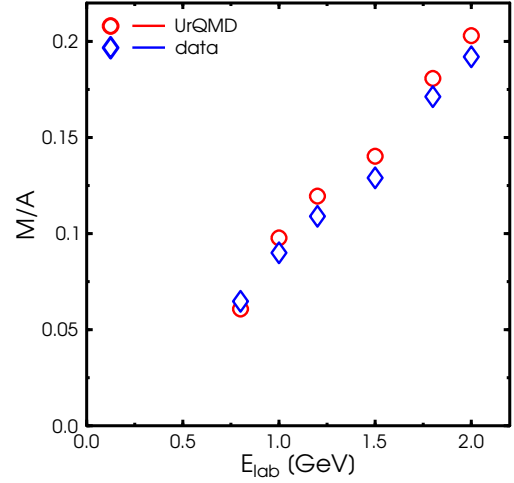,width=.45\textwidth,angle=0}}
\caption{Excitation function of the pion multiplicities for C+C collisions 
from $E_{\rm beam}=0.8$~AGeV to 2.0~AGeV.
Diamonds depict experimental data from the KAOS collaboration \cite{Sturm:2000dm}, squares show the
UrQMD results.} 
\label{excf}
\end{figure}
\begin{figure}[h!]
\vspace*{-1cm}\centerline{\epsfig {file=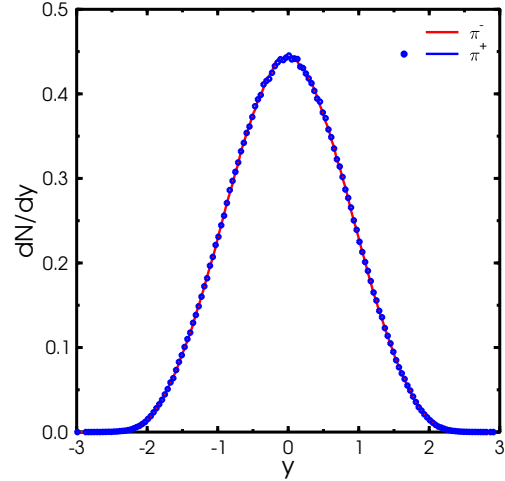,width=.45\textwidth,angle=0}}
\caption{Rapidity distribution of positive and negative Pions for min. bias C+C
reactions at 2~AGeV. The line depicts $\pi^-$, while circles show $\pi^+$.} 
\label{rap_pi}
\end{figure}

For our studies we apply the UrQMD model. It is a non-equilibrium
transport approach based on the covariant propagation of hadrons and strings. All
cross sections are calculated by the
principle of detailed balance or are fitted to data where available. 
The model allows to study the full
space time evolution for all hadrons, resonances and their decay products.
This permits to explore the emission patterns of the resonances 
in detail and to gain insight into the origin of the resonances.
For further details of the model the
reader is referred to \cite{Bass:1998ca,Bleicher:1999xi}.
UrQMD has been successfully applied to study light and heavy ion
reactions at SIS. Detailed comparisons of UrQMD with a large body of
experimental data at SIS energies can be found in \cite{Sturm:2000dm}.

The results shown in this publication are obtained by simulations of
more than 8 $\cdot$ $10^6$ events of min.bias C+C interactions at 2~ AGeV.
The statistical errors in the calculation are therefore small and
will not be shown separately.

To set a baseline for the study of the $\rho^0$ and $\omega$ mesons,
we compare our calculations with the energy dependence of the total
Pion multiplicities in C+C reactions as shown in Fig. \ref{excf}. 
For the small Carbon system, the UrQMD calculation (circles) is in
reasonable agreement with the experimental data from the KAOS
experiment \cite{Sturm:2000dm}.

Let us now turn to the HADES experiment which will soon have first data on 
min.bias C+C reactions at 2~AGeV. In Fig. \ref{rap_pi}
we predict the Pion rapidity spectra for negative and positive
charges. As expected from this isospin symmetric nuclei, the yields of $\pi^+$ and
$\pi^-$ are identical and reach a maximal value of $dN/dy = 0.45$ at $y_{\rm cm}$.

\begin{figure}[t!]
\vspace*{-1cm}\centerline{\epsfig
{file=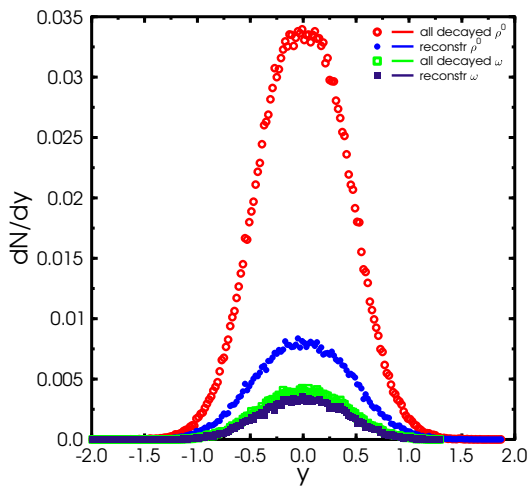,width=.45\textwidth,angle=0}}
\caption{
Rapidity distribution of the decayed mesons for min. bias C+C reactions at
2~AGeV. The open circles/squares show the
$\rho^0_{770}$'s / $\omega$'s which can be reconstructed in the pion channel, full circles
depict all decayed $\rho^0$'s / $\omega$'s.
A strong suppression of in hadron correlations reconstructable resonances
compared to those reconstructable via leptonic decays (indicated as 'all
decayed') is visible for the $\rho^0$ mesons, whereas the spectrum for the $\omega$ meson 
is only weakly altered.
}
\label{rap}
\end{figure}
\begin{figure}[t!]
\vspace*{-1cm}\centerline{\epsfig
{file=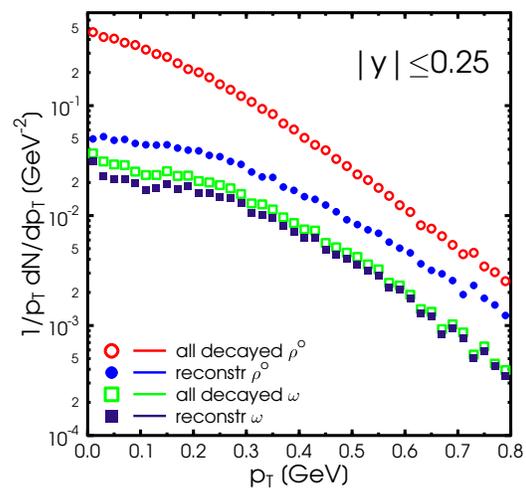,width=.45\textwidth,angle=0}}
\caption{
Transverse momentum distributions of $\rho^0$ mesons (circles) and
$\omega_{770}$ mesons
(squares) for min. bias C+C reactions at 2~AGeV at midrapidity ($|y| \le 
0.25$).
Circles depict $\rho^0$ mesons, squares depict $\omega$ mesons. Open symbols
show all decayed resonances, whereas full symbols show those actually
reconstructable via hadron correlations.  There is a huge modification for the
$\rho^0$ meson spectrum at low
$p_T$, whereas the high $p_T$ part is modified less (factor of 4-5 less
compared to the low $p_T$ part).
There is only a very slight modification visible for the the $\omega$ meson.
}
\label{pt}
\end{figure}
\begin{figure}[h!]
\vspace*{-1cm}\centerline{\epsfig
{file=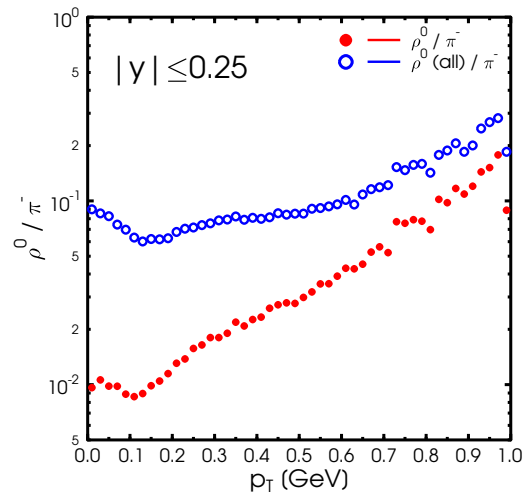,width=.45\textwidth,angle=0}}
\caption{
The $\rho^0/\pi^-$ ratio as a function of transverse momentum for min. bias
C+C reactions at 2~AGeV at midrapidity ($|y| \le
0.25$).
Full circles depict the via hadron correlations reconstructable $\rho^0$'s
over pions, whereas the blue circles depict all decayed $\rho^0$'s over pions.
It is evident that there is less suppression at higher transverse momentum. 
}
\label{ratio}
\end{figure}

After these preliminaries, we focus for the rest of this work on the resonance
production. Experimentally, the identification of resonances proceeds via
the reconstruction of the invariant mass distribution (e.g. of charged pions) for
each event. Then, an invariant mass distribution of mixed events is generated (here the
particle pairs are uncorrelated by definition) and subtracted from the mass 
distribution of the correlated events. As a result one obtains the
mass distributions and yields (after all experimental corrections) of
the resonances by fitting the resulting distribution with a suitable 
function (usually a Breit-Wigner distribution peaked around the pole mass). 
\begin{figure}[t!]
\vspace*{-1cm}\centerline{\epsfig
{file=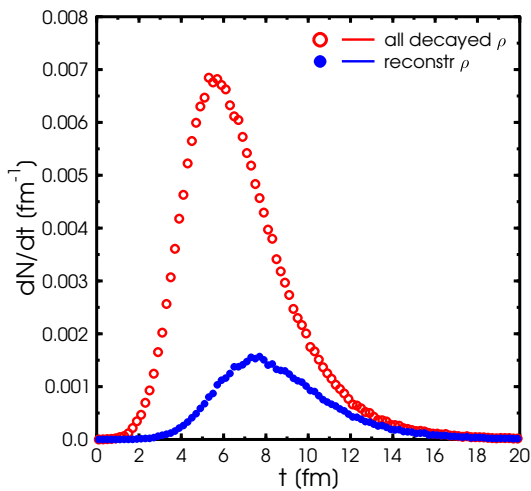,width=.45\textwidth,angle=0}}
\caption{
Temporal distribution of $\rho^0$ resonances for min. bias C+C reactions at
2~AGeV. Open circles depict all decayed
$\rho^0$'s, full circles the via hadron correlations reconstructable ones.
Those visible via pion decay products decay at a very late stage of the
collision, roughly at 7-8 fm/c.
}
\label{time_rho}
\end{figure}

For the model calculation of the resonances, we employ a different 
method to extract the resonances. Here, we follow the decay products of each 
decaying resonance (the daughter particles). If any of the daughter hadrons
rescatters, the signal of this resonance is lost. If the daughter particles
do not rescatter in the further evolution of the system, the resonance is counted
as 'reconstructable'. Note that all decaying resonances are dubbed 
with the term 'all decayed'. These resonances
are reconstructable by an invariant mass analysis of di-leptons (after multiplication with
the respective branching ratio $\Gamma (R\rightarrow e^+e^-)$). 
The advantage of this method is that it allows 
to trace back the origin of each individual resonance to study their spatial and temporal
emission pattern. Even more, it enables one to study the production process of the finally
observed resonance itself, shedding light on the origin of mass modifications.

Figure \ref{rap} depicts the rapidity distributions of the meson resonances for min.bias
C+C interactions at 2 AGeV. The full
symbols display those $\rho^0$ (circles) and $\omega$ (squares) resonances 
which are reconstructable via hadron
correlations. This means that the daughter particles do not interact after
the decay of the resonance. The open symbols show the spectra for all 
decayed  $\rho^0$  and $\omega$ resonances.
This can be interpreted as the spectrum which can be measured via a
di-leptonic decay, not taking into account any interferences of the $\rho$ and
the $\omega$. 
In the case of the $\rho^0$ one observes a drastic reduction of the observable
yield at midrapidity from 3.5\% (in the di-lepton channel) to  0.8\% in the hadronic channel.
In contrast, the $\omega$ meson is only slightly altered when the reconstruction
probabilities in both channels are compared.
This can be traced back to the much longer lifetime of the $\omega$ compared to
the $\rho$. Most of the $\omega$'s will leave the interaction zone before they decay,
thus reducing the possibility of the rescattering of the daughters.

If this interpretation is valid, one expects a strong transverse 
momentum ($p_T = \sqrt{p_x^2+p_y^2}$) dependence of the suppression pattern.
One would expect a larger modification of short lived resonances 
at low transverse momentum. 
The spectrum at higher transverse momenta will only be slightly altered, because high
$p_T$-resonances are more likely to escape from the interaction region before they decay.
As shown in Fig. \ref{pt} the $\rho^0_{770}$ meson is suppressed, especially
at low $p_T$, in line with our expectations from the rescattering picture. Only at very low
transverse momenta, the $\omega_{782}$ is weakly suppressed.
It should be noted that a similar behaviour was also found experimentally for larger systems
at higher energies \cite{Adams:2004ep}. 
\begin{figure}[t!]
\vspace*{-1cm}\centerline{\epsfig
{file=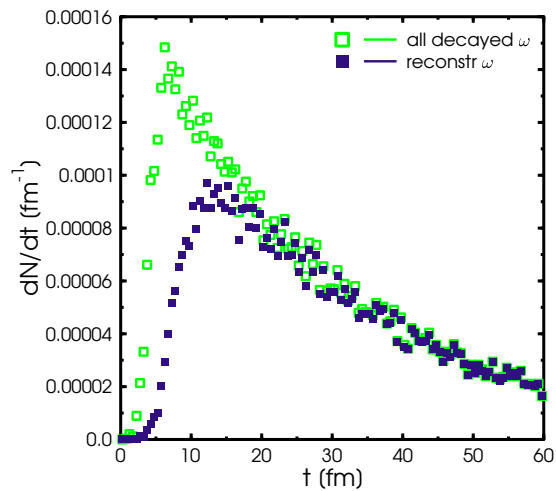,width=.45\textwidth,angle=0}}
\caption{
Temporal distribution of $\omega$ resonances for min. bias C+C reactions at
2~AGeV. Open squares depict all decayed
$\omega$'s, full squares the via hadron correlations reconstructable ones.
Note that the $\omega$ meson decays at a very late stage of a collision due
to its long lifetime. One also observes a long tail of the distribution
because of that. 
}
\label{time_omega}
\end{figure}

Figure \ref{ratio} depicts the $\rho^0/\pi^-$ ratio as a function of $p_T$.
A strong suppression of the $\rho^0/\pi^-$ ratio at low $p_T$ is evident for
the reconstructable resonances, which vanishes at
high $p_T$. This also supports the rescattering scenario, since 
high $p_T$ pions and $\rho^0$'s leave the medium directly and do not 
experience modifications due to
rescattering.

What is the temporal and spatial emission pattern of the resonances observable in the
di-leptonic compared to those in the hadronic channel?
Figures \ref{time_rho} and \ref{time_omega} depict the temporal distributions 
of reconstructable and all decayed $\rho^0$ and $\omega$ mesons.
One observes a shift by 2~fm/c in the peak of the emission times from 5.5~fm/c for
the di-lepton channel compared to the reconstructable $\rho^0$'s  that are emitted at 
later stages  of the collision, around 7.5~fm/c. Overall, the $\omega$'s decay even 
later (notice the different scaling) at 7.5~fm/c (di-lepton channel) and
13 fm/c (pion channel), i.e. outside of the collision zone.
That is the reason why there is nearly no suppression of the $\omega$ meson in
the hadron channel.

\begin{figure}[t!]
\vspace*{-1cm}\centerline{\epsfig
{file=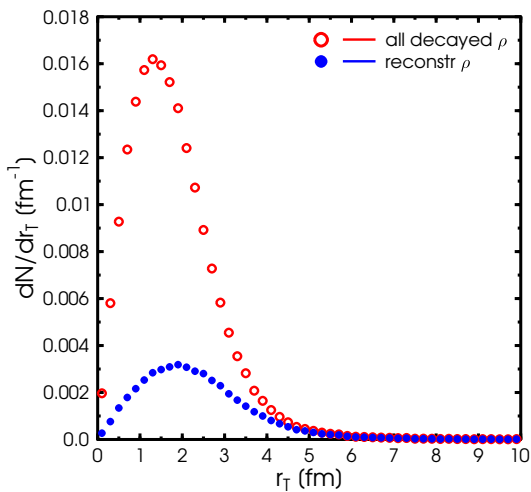,width=.45\textwidth,angle=0}}
\caption{Spatial distribution of $\rho^0$ resonances for min. bias C+C reactions at
2~AGeV. Full symbols depict those resonances whose decay products did not
interact after the decay, whereas open symbols depict all decayed
resonances.
}
\label{rt_rho}
\end{figure}

\begin{figure}[t!]
\vspace*{-1cm}\centerline{\epsfig
{file=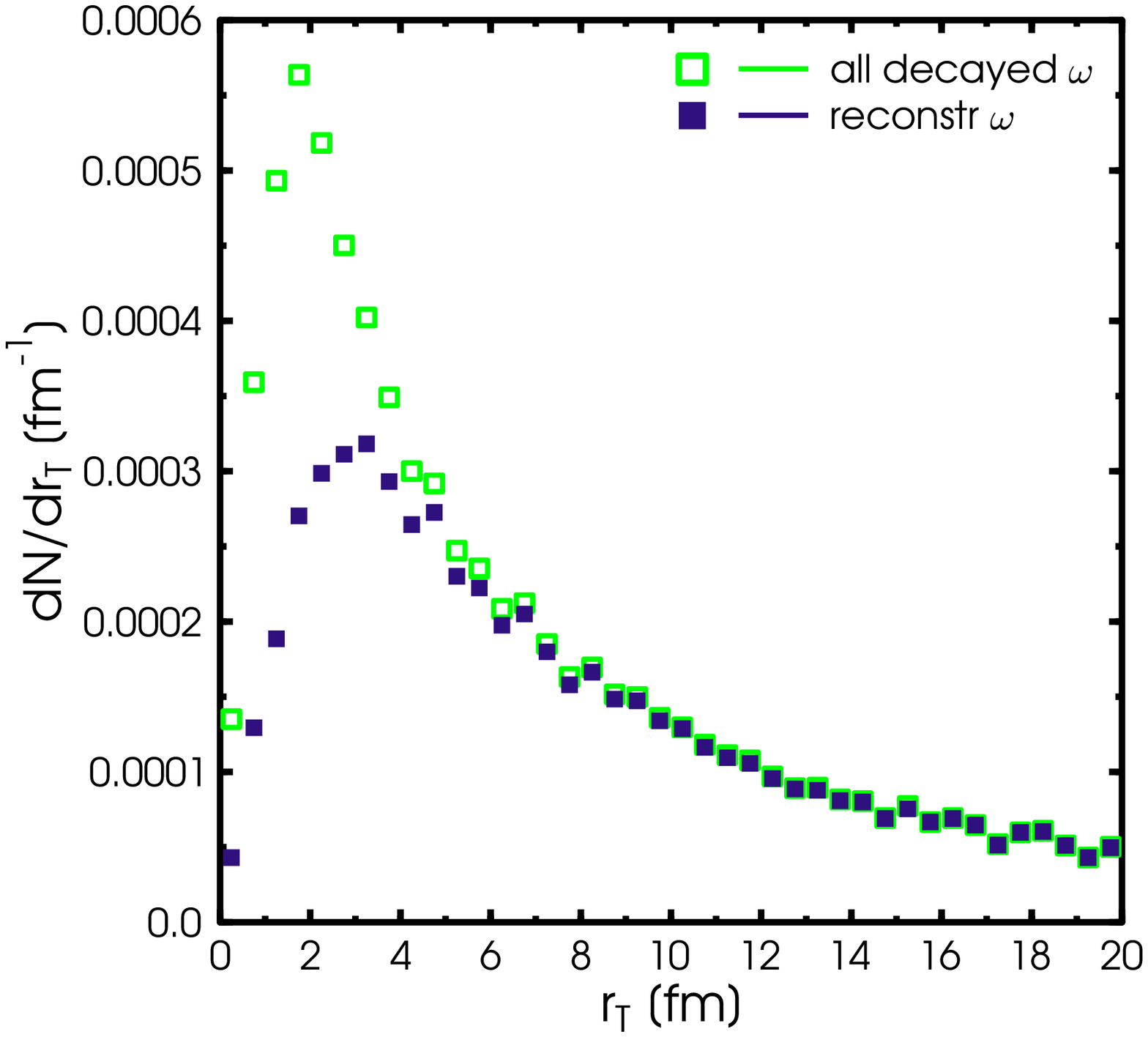,width=.45\textwidth,angle=0}}
\caption{Spatial distribution of $\omega$ resonances for min. bias C+C reactions at
2~AGeV. Full symbols depict those resonances whose decay products did not
interact after the decay, whereas open symbols depict all decayed
resonances.
}
\label{rt_omega}
\end{figure}

Figures \ref{rt_rho} and \ref{rt_omega} depict the transverse distance
$r_T = \sqrt{r_x^2+r_y^2}$
of the meson resonances at the point of their decay. 
Here it is interesting to notice that there is only a rather small difference in the
peak emission radii of the $\rho^0$ resonances observable in the di-lepton spectrum
(maximal emission at $r_T=1.3$~fm)
compared to those in the hadronic correlation with $r_T=1.8$~fm.
Comparing both emission radii to the size of the Carbon nucleus ($r\sim 2.7$ ~fm)
it seems that both reconstruction channels seem to be sensitive to similar 
in-medium modifications. A detailed comparison of the thermal parameters at
the decay point of the resonances will be given in a follow-up work. 

\begin{figure}[h!]
\vspace*{-1cm}\centerline{\epsfig
{file=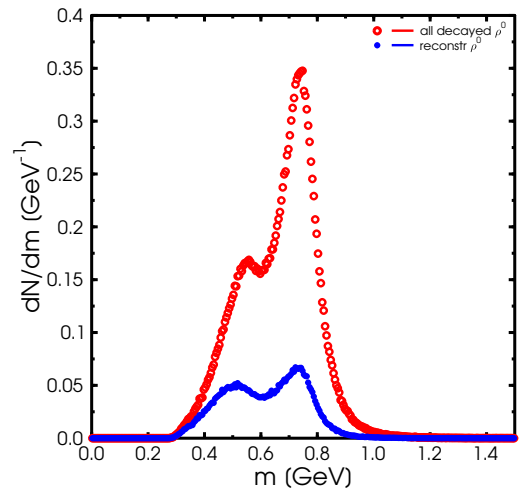,width=.45\textwidth,angle=0}}
\caption{
Mass distribution for $\rho^0$ mesons for min. bias C+C reaction at 2~AGeV. 
The peak around 500~MeV is due to a strong contribution from 
$N^*_{1520} \rightarrow p+\rho$ which amounts to 75\% for masses below 600~MeV. 
}
\label{mass}
\end{figure}

Let us finally discuss, one of the most interesting effects, 
the modification of the mass spectrum of the $\rho^0$ meson as has 
been discussed earlier for example in \cite{Brown:1991kk, Rapp:1999qu, Cassing:1997jz}.
The mass distribution of the $\rho^0$ in min.bias C+C reaction at 2~AGeV is shown 
in Fig. \ref{mass}. Open circles depict all decayed $\rho$'s (this is similar to the
di-leptons invariant mass distribution multiplied by the branching ratio 
and the vector meson dominance factor of $1/m^3$), while the
$\rho$'s reconstructable in $\pi\pi$ correlations are shown as full circles.
In both distributions, one observes a clear double peak structure,
with maxima at the $\rho$ pole mass (770~MeV) and around 500 to 600~MeV.
Usually an enhancement of the $\rho$ spectral function in this 
mass region has been attributed to strong in-medium modifications,
due to finite densities and temperatures.
However, in the present calculation, we do not make explicit use
of any in-medium modification, but only include the coupling of the $\rho$
to pions and baryons via the employed cross sections calculated
from detailed balance.

A detailed analysis shows that the low mass peak is due to the
decay chain $N^*_{1520} \rightarrow p+\rho^0$ which contributes to 75\%
to the reconstructable $\rho$ mass spectrum below 600~MeV.
Without in-medium modifications, this decay process restricts the mass of the
$\rho$ to $m_\rho\le m_{N^*_{1520}}-m_p\sim 580$~MeV and thus feeds strongly
into the low invariant mass region of the $\rho$.
Above 600~MeV, $\rho$'s are mostly produced from $\pi\pi\rightarrow\rho$.
It seems that a dramatic modification of the $\rho$ spectral function
is mostly due to the decay kinematics of the production channel of the
$\rho$. However, on top of these  kinematic effects additional modifications
of the $\rho$ mass spectrum might occur.

In summary, we have explored $\rho$ and $\omega$ production in C+C interactions
at 2~AGeV. We have predicted the yields and spectra of these meson resonances
when reconstructed in  ''di-lepton-like'' and hadronic channel.
The present calculations show a strong difference (factor 5) between the
yields observable in the di-lepton and hadron channel.
This can be understood due to a rescattering of the resonance 
daughter particles, shifting them out of the $\rho$ peak.
At midrapidity $8\cdot 10^{-3}$ $\rho^0$'s per event can be reconstructed
from the pion correlations.  
We predict a strong transverse momentum dependence of the $\rho$ suppression
pattern leading to an apparent heat-up of the $\rho$'s observed in the hadronic
channel compared to the di-lepton channel.
Finally, we have pointed out that the mass spectrum of the reconstructable $\rho$'s
shows a strong double peak structure. 
This second peak around an invariant mass of 500~MeV is due to $\rho$'s
from the decay of the $N^*(1520)$ which feeds directly in to the 
$\rho$ mass region below 580~MeV.
Our prediction are a complimentary approach to the di-lepton measurements 
underway at HADES/GSI. The reconstruction of resonances in the hadronic channel 
yields additional information on the later stages of the reaction and
is of special interest for direct tests of hadronic transport 
models without involving any di-lepton ''after burners''.

\section*{Acknowledgements}

The authors thank  GSI and BMBF for support. 
We thank Diana Schumacher and Christian Sturm for detailed discussions about the KAOS data.
Fruitful discussions with  Horst St\"ocker, Peter Zumbruch  and 
Joachim Stroth are greatfully acknowledged.
The computational resources have been provided by the 
Center for Scientific Computing in Frankfurt.



\begin{thebibliography}{00}





\bibitem{Markert:2004xx}
C.~Markert  [STAR Collaboration],
J.\ Phys.\ G {\bf 30} (2004) S1313
[arXiv:nucl-ex/0404003].

\bibitem{Markert:2003rw}
C.~Markert  [STAR Collaboration],
arXiv:nucl-ex/0308029.

\bibitem{Markert:2002xi}
C.~Markert  [STAR Collaboration],
J.\ Phys.\ G {\bf 28} (2002) 1753
[arXiv:nucl-ex/0308028].

\bibitem{Fachini:2003dx}
P.~Fachini  [STAR Collaboration],
J.\ Phys.\ G {\bf 30} (2004) S565
[arXiv:nucl-ex/0305034].

\bibitem{Fachini:2003mc}
P.~Fachini,
Acta Phys.\ Polon.\ B {\bf 35} (2004) 183
[arXiv:nucl-ex/0311023].

\bibitem{Fachini:2004jx}
P.~Fachini,
J.\ Phys.\ G {\bf 30} (2004) S735
[arXiv:nucl-ex/0403026].

\bibitem{Adams:2004ep}
  J.~Adams {\it et al.}  [STAR Collaboration],
  arXiv:nucl-ex/0412019.


\bibitem{Adler:2004zn}
S.~S.~Adler {\it et al.}  [PHENIX Collaboration],
arXiv:nucl-ex/0409015.

\bibitem{Johnson:1999fv}
S.~C.~Johnson, B.~V.~Jacak and A.~Drees,
Eur.\ Phys.\ J.\ C {\bf 18} (2001) 645
[arXiv:nucl-th/9909075].


\bibitem{Bleicher:2002dm}
M.~Bleicher and J.~Aichelin,
Phys.\ Lett.\ B {\bf 530} (2002) 81
[arXiv:hep-ph/0201123].

\bibitem{Torrieri:2002jp}
  G.~Torrieri and J.~Rafelski,
  Phys.\ Rev.\ C {\bf 68} (2003) 034912
  [arXiv:nucl-th/0212091].

\bibitem{Bleicher:2002rx}
  M.~Bleicher,
  Nucl.\ Phys.\ A {\bf 715} (2003) 85
  [arXiv:hep-ph/0212378].

\bibitem{Bleicher:2003ij}
  M.~Bleicher and H.~Stocker,
  J.\ Phys.\ G {\bf 30} (2004) S111
  [arXiv:hep-ph/0312278].

\bibitem{Torrieri:2004zz}
  G.~Torrieri, S.~Steinke, W.~Broniowski, W.~Florkowski, J.~Letessier and J.~Rafelski,
  arXiv:nucl-th/0404083.

\bibitem{Vogel:2005qr}
  S.~Vogel and M.~Bleicher,
  arXiv:nucl-th/0505027.


\bibitem{Soff:2000ae}
  S.~Soff {\it et al.},
  J.\ Phys.\ G {\bf 27} (2001) 449
  [arXiv:nucl-th/0010103].


\bibitem{Torrieri:2001ue}
G.~Torrieri and J.~Rafelski,
Phys.\ Lett.\ B {\bf 509} (2001) 239
[arXiv:hep-ph/0103149].

\bibitem{Torrieri:2001tg}
  G.~Torrieri and J.~Rafelski,
  J.\ Phys.\ G {\bf 28} (2002) 1911
  [arXiv:hep-ph/0112195].


\bibitem{Braun-Munzinger:1995bp}
  P.~Braun-Munzinger, J.~Stachel, J.~P.~Wessels and N.~Xu,
  Phys.\ Lett.\ B {\bf 365} (1996) 1
  [arXiv:nucl-th/9508020].
  
\bibitem{Braun-Munzinger:2001ip}
  P.~Braun-Munzinger, D.~Magestro, K.~Redlich and J.~Stachel,
  Phys.\ Lett.\ B {\bf 518} (2001) 41
  [arXiv:hep-ph/0105229].


\bibitem{kaischweda}
  K.~Schweda,
  Talk given at the Quark Matter 2004
  



\bibitem{Bass:1998ca}
S.~A.~Bass {\it et al.},
Prog.\ Part.\ Nucl.\ Phys.\  {\bf 41} (1998) 225
[arXiv:nucl-th/9803035].

\bibitem{Bleicher:1999xi}
M.~Bleicher {\it et al.},
J.\ Phys.\ G {\bf 25} (1999) 1859
[arXiv:hep-ph/9909407].


\bibitem{Sturm:2000dm}
  C.~Sturm {\it et al.}  [KAOS Collaboration],
  Phys.\ Rev.\ Lett.\  {\bf 86} (2001) 39
  [arXiv:nucl-ex/0011001].


\bibitem{Brown:1991kk}
  G.~E.~Brown and M.~Rho,
  Phys.\ Rev.\ Lett.\  {\bf 66} (1991) 2720.

\bibitem{Rapp:1999qu}
  R.~Rapp and C.~Gale,
  Phys.\ Rev.\ C {\bf 60} (1999) 024903
  [arXiv:hep-ph/9902268].

\bibitem{Cassing:1997jz}
  W.~Cassing, E.~L.~Bratkovskaya, R.~Rapp and J.~Wambach,
  Phys.\ Rev.\ C {\bf 57} (1998) 916
  [arXiv:nucl-th/9708020].
















\end{thebibliography}
\end{document}